\documentstyle[12pt,epsf]{article}
\newcommand{\wt}{\widetilde}

\def\diag{\mathop{\rm diag}\nolimits}

\begin{document}

\begin{titlepage}
\title{\hfill\parbox{4cm}
       {\normalsize MIT-CTP-2939\\YITP-00-3\\{\tt hep-th/0001105}\\January 2000}\\
       \vspace{2cm}
       T-duality of non-commutative gauge theories
       \vspace{1cm}}
\author{Yosuke Imamura\thanks{E-mail: \tt imamura@ralph2.mit.edu}%
\\[20pt]
{\it Center for Theoretical Physics,}\\
{\it Massachusetts Institute of Technology,}\\
{\it Cambridge, MA  02139-4307, USA}
\\[7pt]
{\it and}
\\[7pt]
{\it Yukawa Institute for Theoretical Physics,}\\
{\it Kyoto University, Kyoto 606-8502, Japan}
}
\date{}

\maketitle
\thispagestyle{empty}

\vspace{0cm}

\begin{abstract}
\normalsize
Via T-duality a theory of open strings on a D1-brane wrapped along a cycle of slanted
torus is described by a $U(1)$ gauge theory on a D2-brane in the $B$-field
background.
It is also known that there is another dual description of the D1-brane configuration
by a non-commutative gauge theory on a D2-brane.
Therefore, these two gauge theories on D2-branes are equivalent.
Recently, the existence of a continuous set
of equivalent gauge theories including these two was suggested.
We give a dual D1-brane configuration for each theory in this set,
and show that the relation among parameters for equivalent gauge theories can be
easily reproduced by rotation of the D1-brane configuration.
We also discuss a relation between this duality and Morita equivalence.
\end{abstract}

%\vspace{3cm}

\vfill

\noindent
%PACS codes : 11.25.Sq, 11.25.-w, 11.15.Pg.\\
%Keywords   : AdS, baryon, brane configuration, supersymmetry.

\end{titlepage}
%\tableofcontents
%%%%%%%%%%%%%%%%%%%%%%%%%%%%%%%%%%%%%%%%%%%%%%%%%%%%%%%%%%%%%%%%%%%%%%%
%%%%%%%%%%%%%%%%%%%%%%%%%%%%%%%%%%%%%%%%%%%%%%%%%%%%%%%%%%%%%%%%%%
%%%%%%%%%%%%%%%%%%%%%%%%%%%%%%%%%%%%%%%%%%%%%%%%%%%%%%%%%%%%%%%%%%%%%%%
\section{Introduction}
Non-commutative gauge theories have attracted much attention
since the existence of the limit realizing non-commutative
gauge theories as theories on D-branes was found.\cite{CDS,dh}
It is known that the non-commutativity on D-branes
is due to the background $B$-field.
By quantizing open strings in the non-vanishing $B$-field background,
we obtain non-vanishing commutators of coordinates of the end points
of the strings on D-branes.
\cite{chuho,fz,kimoh}

Let us restrict our attention to two-dimensional
commutative and non-commutative $U(1)$ gauge theories on
D2-branes wrapped around a rectangular torus for simplicity.
The easiest way to see the non-commutativity is T-dualizing the configuration.
If the background $B$-field is non-zero, the D2-brane is
transformed into
a D1-brane wrapped around one cycle of a slanted torus.
In \cite{dh}, it is noticed that two end points of an open string
on the D1-brane
split due to the slant of the torus,
and interaction among such open strings is
described by non-commutative geometry.
In the context of open strings on a D2-brane,
the splitting of the endpoints is
a result of the interaction of the string worldsheet
with the $B$-field.

Recently, Pioline and Schwarz\cite{morita3} and
Seiberg and Witten\cite{sw} suggested the
existence of a set of equivalent theories parameterized by
the non-commutativity parameter $\theta$.
This set contains
a commutative gauge theory with the background $B$ field,
a non-commutative gauge theory with a vanishing background $2$-form field
and other infinitely many theories with a non-zero background $2$-form field and $\theta$.
In this paper, we will refer to the background $2$ form field in gauge theories
as $\phi$-field because when $\theta\neq0$ the background field
is not identified with the $B$-field in the background spacetime.
The non-commutative Born-Infeld action depends only on the sum $f+\phi$,
where $f$ is a field strength of a gauge field on the brane.

Because it is known that
in the two special cases, $\phi=0$ case and $\theta=0$ case,
the theories are regarded as T-duals of the same D1-brane configuration
on a slanted torus,
it is natural to ask whether such a T-dual picture exists
in the case of general $\phi$ and $\theta$.
In this paper, we suggest the T-dual configuration of D1-brane
for non-commutative $U(1)$ gauge theories with general $\phi$,
$f$ and $\theta$,
and show that the relation of parameters for equivalent gauge theories
can be easily obtained by simple rotation of the D1-brane configuration.

Furthermore, we show that Morita equivalence is generated by combining this T-duality
and the modular transformation of the dual D1-brane configuration.

%%%%%%%%%%%%%%%%%%%%%%%%%%%%%%%%%%%%%%%%%%%%%%%%%%%%%%%%%%%%%%%%%%%%%%%
\section{Ordinary T-duality}
First, let us review the T-duality between ordinary (commutative) gauge theories
on a D2-brane and D1-brane configurations.
For simplicity, we will consider only rectangular D2-branes.
Let $x$ and $y$ be the two coordinates on the D2-brane
and $L_x$ and $L_y$ be the lengths of sides along $x$ and $y$ direction respectively.
Because we will carry out T-duality along the $y$ direction,
the two sides of the rectangle with constant $y$ should be identified.
The other two sides do not have to be identified and we can put $L_x$ infinite.
However, to see change of the metric along $x$ direction,
it is convenient to pick out a finite part with finite length $L_x$.
Thus the background manifold is not a torus but a cylinder.

We will use the following two metrics on the D2-brane.
\begin{equation}
0\leq x\leq L_x,\quad
0\leq y\leq L_y,\quad
G_{ij}=\diag(1,1),
\label{metric1}
\end{equation}
\begin{equation}
0\leq x\leq 1,\quad
0\leq y\leq 1,\quad
G_{ij}=\diag(L_x^2,L_y^2).
\label{metric2}
\end{equation}
Although the metric (\ref{metric2}) is used in much of the literature,
we will mainly use the metric (\ref{metric1}) because in this metric
it is easy to intuitively imagine the meanings of physical values.
Let $\phi$ and $f$ denote the $\phi$-field and the field strength of the gauge field
in the metric (\ref{metric1}).
The total flux $\Phi$ and $F$ of these fields in the whole rectangle are
\begin{equation}
\Phi=L_xL_y\phi,\quad
F=L_xL_yf,
\label{bflux}
\end{equation}
and these are also the values of the fields in the metric (\ref{metric2}).

Let us carry out the T-duality transformation along $y$ direction,
and call the new compactification period $\wt L_y$.
The relation between $L_y$ and $\wt L_y$ is given by
\begin{equation}
\wt L_y=\frac{2\pi}{TL_y},
\label{dualradii}
\end{equation}
where $T$ is the string tension.
\begin{figure}[htb]
\centerline{
\epsfbox{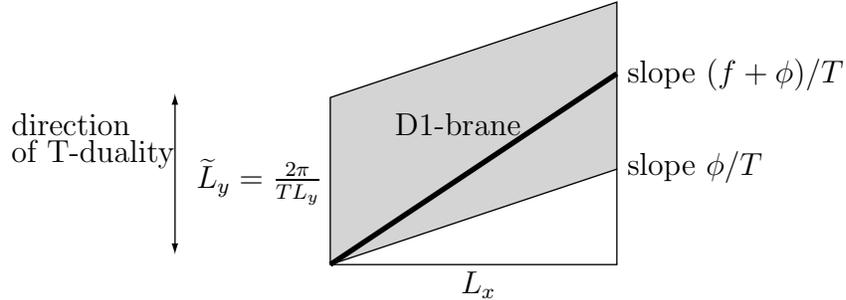}
\put(-60,-10){$L_x$}
\put(3,35){slope $\phi/T$}
\put(3,70){slope $(f+\phi)/T$}
\put(-160,30){$\wt L_y=\frac{2\pi}{TL_y}$}
\put(-230,50){direction}
\put(-230,40){of T-duality}
\put(-85,50){D1-brane}
}
\caption{The dual configuration of the $U(1)$ commutative gauge theory
with the background $2$-form field $\phi$ and
the gauge flux $f$
on a rectangle with size $L_x\times L_y$.}
\label{tdual1.eps}
\end{figure}
By this duality, the flux $\Phi$ is transformed into the twist of the cylinder.
Namely, if we go $L_x$ along $x$ direction,
the ${\bf S}^1$ along $y$ direction is shifted by the angle $\Phi$.
Therefore, the slope of the base of the parallelogram
obtained by cutting the cylinder along $y=0$ line is given by
(Fig.\ref{tdual1.eps})
\begin{equation}
\mbox{slope of the base}=\frac{\phi}{T}.
\label{baselineslope}
\end{equation}
The Wilson line along $y$ direction on the D2-brane
is transformed into the $y$ coordinate of the D1-brane.
In the metric (\ref{metric1}),
gauge field $a_y$ and the coordinate $y$ are related by
\begin{equation}
2\pi\frac{y}{\wt L_y}=L_ya_y.
\end{equation}
(Because we are considering the T-duality along $y$
direction, the background gauge field should be constant along $y$ direction.)
Differentiating this relation with respect to $x$,
we obtain
\begin{equation}
\frac{dy}{dx}=\frac{f}{T},
\end{equation}
where $f=\partial_xa_y$.%
Taking account of the slope of the base of the parallelogram,
the net slope of the D1-brane is represented as follows.
\begin{equation}
\mbox{slope of the D1-brane}=\frac{f+\phi}{T}.
\label{d1braneslope}
\end{equation}
Although the relations (\ref{baselineslope}) and (\ref{d1braneslope}) still hold
if $\phi$ and $f$ depend on the coordinate $x$,
we assume they are constant in what follows.

%%%%%%%%%%%%%%%%%%%%%%%%%%%%%%%%%%%%%%%%%%%%%%%%%%%%%%%%%%%%%%%%
\section{T-duality of non-commutative gauge theory ($\phi=f=0$ case)}
There is another dual description of
a theory of open strings on the D1-brane on the slanted cylinder.
As shown in \cite{dh}, it is equivalent
to a non-commutative gauge theory.
%Formally, this theory can be regarded as a result of `T-duality' along the
%direction vertical to the D1-brane.

\begin{figure}[htb]
\centerline{
\epsfbox{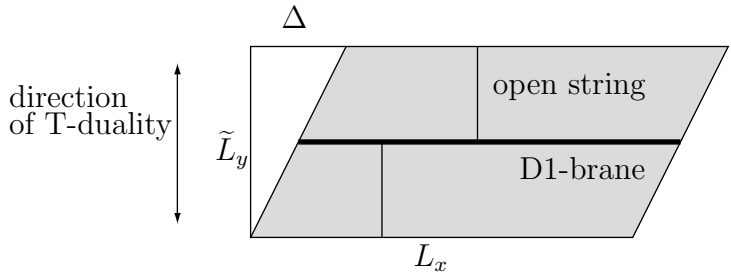}
\put(-195,30){$\wt L_y$}
\put(-120,-10){$L_x$}
\put(-170,80){$\Delta$}
\put(-273,50){direction}
\put(-273,40){of T-duality}
\put(-80,23){D1-brane}
\put(-90,55){open string}
}
\caption{The dual configuration of the $U(1)$ non-commutative gauge theory
with the non-commutativity parameter $\theta=\Delta/T\wt L_y$
on a rectangle with size $L_x\times L_y$.}
\label{tdual2.eps}
\end{figure}

Let us consider the case of vanishing gauge field strength.
In this case, the D1-brane is
parallel to the base of the parallelogram.(Fig.\ref{tdual2.eps})
Let $L_x$ and $\wt L_y$ be the length of the base and
the height of the parallelogram,
respectively.
(These definitions are different from those in the previous section.)
By the duality, this is transformed into non-commutative gauge theory on a
rectangular cylinder.
The width of the rectangle is $L_x$, and
the compactification period $L_y$ along $y$ direction
is determined by equating the energy $T\wt L_y$ of
a wrapped open string and Kaluza-Klein momentum $2\pi/L_y$ in the gauge theory.
As a result, the relation of $L_y$ and $\wt L_y$ is the same with that
for the ordinary T-duality (\ref{dualradii}).

This duality is very similar to the ordinary T-duality,
and so, we will refer to it as T-duality, too.
The different point between these two T-dualities is that
we use closed strings in the ordinary T-duality,
while we use open strings on D1-brane in the new T-duality.
Because open strings are always at right angle to the D1-brane,
their direction and length are different from those of
closed strings.
This causes the difference of
the size of the cylinder of the dual gauge theories.

We can also reproduce ordinary T-duality by using open strings.
In this case, in order to obtain the relation between $L_y$ and $\wt L_y$,
we should use open strings parallel to the left and right sides of the parallelogram.
They are not perpendicular to the D1-brane and
are the solutions of equation of motion only when
we neglect the non-diagonal part of the metric of the slanted cylinder.
This correspond to neglect of the background $B$-field in the
D2-brane picture.
In fact, it is known that, when we carry out the path integral of open strings,
if we do not include the $B$-field term in the kinetic term we obtain
an ordinary gauge theory as a low energy effective theory,
and if we regard it as a part of the kinetic term
we obtain a non-commutative one.\cite{lee}
Therefore, we can say that a gauge theory which we obtain by the T-duality
from the D1-brane configuration depends on the direction of the open strings
which we use.
We refer to this direction as `direction of T-duality.'

The shift between the upper base and the lower base causes the
splitting of the end points of a wrapped open string on the D1-brane.
Let $\Delta$ denote the amount of the shift.
The length of an open string with winding number $n$ is $n\wt L_y$
and the splitting of the end points is $n\Delta$.
This string corresponds to an open string on D2-brane
moving along $y$ direction with momentum $nT\wt L_y$
whose end points split along $x$ by the distance $n\Delta$.
This splitting of end points is the origin of the non-commutativity
of the gauge theory and the parameter of the non-commutativity $\theta$
is given as a ratio of the momentum and the splitting.
\begin{equation}
\theta=\frac{\Delta}{T\wt L_y}.
\label{thetais}
\end{equation}
The parameter $\theta$ defined by (\ref{thetais}) is a value in the metric (\ref{metric1}).
Since this has dimension of length$^2$,
the value $\Theta$ in the metric (\ref{metric2}) is given by
\begin{equation}
\Theta=\frac{1}{L_xL_y}\theta.
\label{thetas}
\end{equation}

We can show that interaction among open strings on the D1-brane
is described with the $*$-product defined with the parameter $\theta$
in the dual gauge theory.\cite{dh}
Let $\phi_n(x)$ denote a field of open strings
with winding number $n$ and endpoints at $x\pm n\Delta/2$.
Because open strings interact at their end points,
the interaction is
represented by using the following non-local product.
\begin{equation}
\phi^1_n(x)=\sum_k\phi^2_{n-k}\left(x-\Delta\frac{n}{2}\right)
                   \phi^3_k\left(x+\Delta\frac{n-k}{2}\right)
\label{interaction}
\end{equation}
The dual field $\phi(x,y)$ on the D2-brane
is obtained from this field by the following
Fourier transformation.
\begin{equation}
\phi(x,y)=\sum_ne^{2\pi iny/L_y}\phi_n(x).
\label{fourier}
\end{equation}
Using this field, the product (\ref{interaction})
is rewritten as the following $*$-product.
\begin{eqnarray}
\phi^1(x,y)&=&\phi^2(x,y)*\phi^3(x,y)\nonumber\\
&\equiv&
\exp\frac{i\theta}{2}
(\partial_{x_2}\partial_{y_3}-\partial_{y_2}\partial_{x_3})
\phi^2(x_2,y_2)
\phi^3(x_3,y_3)|_{x_2=x_3=x,y_2=y_3=y}
\label{pxyprod}
\end{eqnarray}
%%%%%%%%%%%%%%%%%%%%%%%%%%%%%%%%%%%%%%%%%%%%%%%%%%%%%%%%%%%%%%%%%%%%%%%%%%%%
%%%%%%%%%%%%%%%%%%%%%%%%%%%%%%%%%%%%%%%%%%%%%%%%%%%%%%%%%%%%%%%%%%

\section{T-duality with general parameters}
\begin{figure}[htb]
\centerline{
\epsfbox{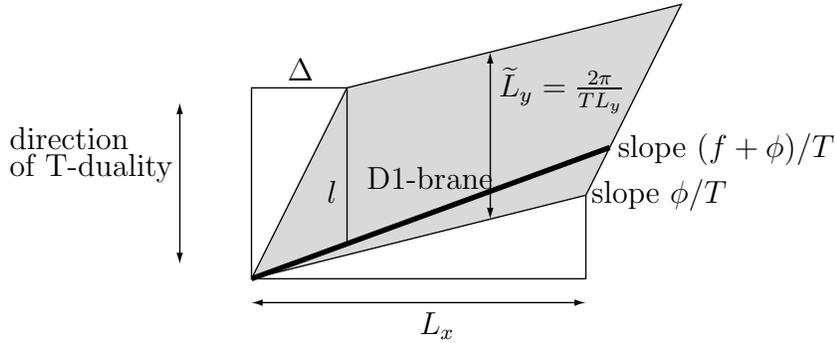}
\put(-100,-10){$L_x$}
\put(-35,40){slope $\phi/T$}
\put(-25,57){slope $(f+\phi)/T$}
\put(-150,86){$\Delta$}
\put(-135,40){$l$}
\put(-70,80){$\wt L_y=\frac{2\pi}{TL_y}$}
\put(-255,60){direction}
\put(-255,50){of T-duality}
\put(-120,45){D1-brane}
}
\caption{The dual configuration of the $U(1)$ non-commutative gauge theory
with general values of $f$, $\phi$ and $\theta$
on a rectangle with size $L_x\times L_y$.}
\label{phi2.eps}
\end{figure}

In this section, we generalize the duality.
Let us consider a non-commutative $U(1)$ gauge theory
on a rectangular cylinder with size $L_x\times L_y$,
and let $\phi$, $f$ and $\theta$ denote the background $2$-form field, gauge field strength
and non-commutativity parameter in the metric (\ref{metric1}), respectively.
What D1-brane configuration is dual to this theory?
Now, we suggest that the relations
(\ref{dualradii}), (\ref{baselineslope}), (\ref{d1braneslope})
and (\ref{thetais}) still hold in the general case
if we define the parameters $L_x$, $\wt L_y$ and $\Delta$
as shown in Fig.\ref{phi2.eps}.
(For $\phi=0$ case, the relation (\ref{d1braneslope}) was already obtained in \cite{aaj}.)
We will not prove this statement.
Instead, we would like to give two comments which make it seem reasonable.

The first one is about the compactification period $L_y$.
Because the length of an open string along the direction of T-duality%
\footnote{This string is a solution of equation of motion
if we take account of a particular portion of the non-diagonal element of the metric of the cylinder.
In a low energy effective field theory obtained by
quantizing the open strings,
the rest would appear as a background $\phi$-field in the gauge theory.
}
with winding number one is
not $\wt L_y$ but $l=\wt L_y-\Delta f/T$ in the present case
as shown in Fig.\ref{phi2.eps},
the Kaluza-Klein momentum in the dual gauge theory should be quantized
with a unit $lT$.
This unit is different from ordinary one $2\pi/L_y$.
This difference is interpreted as follows.
In a non-commutative gauge theory, even if the gauge group is $U(1)$,
the gauge field couples to an adjoint matter field $\psi$
via the non-commutativity.
For example, the covariant derivative $\nabla_y\psi$ is not just a
partial derivative.
\begin{equation}
\nabla_y\psi=\partial_y\psi+ia_y*\psi-i\psi*a_y.
\end{equation}
If the field strength $f=\partial_xa_y$ is constant,
we obtain the following relation.
\begin{equation}
\nabla_y\psi=(1-f\theta)\partial_y\psi.
\end{equation}
Because the factor $1-f\theta$ in the right hand side can be removed by the
rescaling of the coordinate $y$,
the interaction with the background gauge field effectively
changes the metric,
and the effective period $L_y^{\rm eff}$ along $y$ direction is
\begin{equation}
L_y^{\rm eff}=\frac{L_y}{1-f\theta}.
\end{equation}
The unit of the Kaluza-Klein momentum $lT$ is
related to this effective period in the usual way.

The other comment is on the quantization of the gauge flux.
If we identify the left side and the right side of the parallelogram,
the D1-brane has to go through the corresponding points on the left and right sides.
As a result, the slope of the D1-brane is quantized.
This correspond to the flux quantization on the D2-brane.
Due to the slant of the sides
the quantized slopes are {\em not} multiples of some constant.
It is given by
\begin{equation}
\frac{f}{T}=\frac{\wt L_yn}{L_x+\Delta n},\quad
n\in{\bf Z}.
\label{qslope}
\end{equation}
This relation is reproduced in the framework of the non-commutative gauge theory
as follows.
For the purpose of T-duality,
the gauge field should not depend on the coordinate $y$,
and if the flux is constant the gauge field is given by
\begin{equation}
a_y(x)=fx.
\end{equation}
The gauge field on the boundary $a_y(0)$ and $a_y(L_x)$ should be
equivalent up to a gauge transformation.
\begin{equation}
a_y(0)=-iU(y)*\partial_yU^{-1}(y)
       +U(y)*a_y(L_x)*U^{-1}(y).
\label{gaugeequiv}
\end{equation}
To keep the gauge field independent of the coordinate $y$,
$U(y)$ should be the following function.
\begin{equation}
U(y)=e^{2\pi iny/L_y},\quad
n\in{\bf Z}.
\end{equation}
The quantization of $n$ is due to the periodicity along $y$ direction.
Using the definition of the $*$-product (\ref{pxyprod}), we can show an identity
\begin{equation}
e^{iky}*u(x)*e^{-iky}=u(x+k\theta),
\end{equation}
for an arbitrary function $u(x)$,
and (\ref{gaugeequiv}) is rewritten as
\begin{equation}
a_y(0)=-\frac{2\pi n}{L_y}+a_y\left(L_x+2\pi n\frac{\theta}{L_y}\right).
\end{equation}
This equation restricts the value of $f$ into the following value
\begin{equation}
f=\frac{2\pi n/L_y}{L_x+2\pi n\theta/L_y},
\end{equation}
and this is equivalent to the relation (\ref{qslope}).

\begin{figure}[htb]
\centerline{
\epsfbox{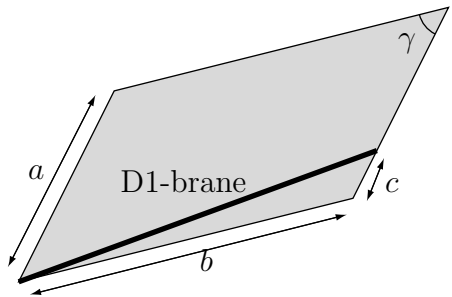}
\put(-160,45){$a$}
\put(-95,10){$b$}
\put(-20,95){$\gamma$}
\put(-25,40){$c$}
\put(-125,40){D1-brane}
}
\caption{The definition of placement-independent variables $a$, $b$, $c$ and $\gamma$.}
\label{para.eps}
\end{figure}

Recently, it was suggest that
if a matrix $M^{ij}$ defined by the following equation (\ref{mis})
for two non-commutative gauge theories with different parameters are the same,
these theories are equivalent.\cite{morita3,sw}
The matrix $M^{ij}$ is defined by
\begin{equation}
M^{ij}=\frac{1}{G_{ij}-\Phi_{ij}/T}+T\Theta^{ij},
\label{mis}
\end{equation}
where $G_{ij}$ is the metric in (\ref{metric2})
and $\Phi_{ij}$ and $\Theta^{ij}$ are the following antisymmetric matrices representing
the background $\phi$-field and the non-commutativity.
(Our definition of $\Phi$ is different from that in \cite{morita3,sw} by the signature.)
\begin{equation}
\Phi_{ij}=\left(\begin{array}{cc}
             & -\Phi \\
             \Phi \end{array}\right),\quad
\Theta^{ij}=\left(\begin{array}{cc}
             & -\Theta \\
             \Theta \end{array}\right).
\end{equation}
$\Phi$ and $\Theta$ are defined with metric (\ref{metric2})
and are related to $\phi$ and $\theta$ in metric (\ref{metric1})
by (\ref{bflux}) and (\ref{thetas}).
Substituting these into (\ref{mis}),
we obtain the following elements of the matrix $M^{ij}$.
\begin{equation}
M^{ij}
=\frac{1}{L_xL_y(1+\phi^2/T^2)}\left(\begin{array}{cc}
         L_y/L_x & -\phi/T-(1+\phi^2/T^2)T\theta \\
         \phi/T+(1+\phi^2/T^2)T\theta & L_x/L_y
         \end{array}\right).
\end{equation}
These elements can be rewritten as
\begin{equation}
M^{ij}=\left(\begin{array}{cc}
          1/b^2 & -(Ta/2\pi b)\cos\gamma \\
          (Ta/2\pi b)\cos\gamma & (Ta/2\pi)^2\sin^2\gamma
         \end{array}\right),
\end{equation}
where $a$, $b$ and $\gamma$ are
the lengths of two sides and the angle of a corner(Fig.\ref{para.eps}).
These parameters are independent of the placement of the parallelogram
and invariant under its rotation.
Therefore, in the dual picture,
all the parallelograms corresponding to equivalent non-commutative gauge theories
are congruent with each other,
and the change of parameters among the equivalent theories is regarded as simple rotation of the
parallelogram.

Next, let us consider the gauge field.
In \cite{sw}, the relation among gauge fields
of the equivalent non-commutative gauge theories
with different $\theta$ is given as a solution of
a certain differential equation.
The equation can be solved in the case of constant gauge field strength,
and the solution is
\begin{equation}
\frac{1}{F}-\Theta=\frac{1}{F_0},
\label{equiv}
\end{equation}
where $F$ is gauge field strength for a theory with non-commutativity $\Theta$
and $F_0$ is that for $\Theta=0$.
This equation implies that
all equivalent gauge theories have a common value of $1/F-\Theta$.
In the dual D1-brane configuration, this value is represented by the placement-independent variables
as follows.
\begin{equation}
\frac{1}{F}-\Theta=\frac{a}{2\pi c},
\end{equation}
where $c$ is defined in Fig.\ref{para.eps}.
Therefore, the equivalence relation of gauge fields is also reproduced by rotation of
the parallelogram.

\section{Morita equivalence}
There is another kind of equivalence among non-commutative gauge theories
which is called Morita equivalence.\cite{morita1,morita2,morita3}
Morita equivalence is `T-duality' among non-commutative gauge theories.
We stress that the duality which we discussed in the previous section
is not included in Morita equivalence.
In two-dimensional case,
Morita equivalence is regarded as a duality among D2-brane configurations,
while we have discussed the duality between D2-branes and D1-branes.
These two duality, however, is intimately related.

In the two-dimensional case, Morita equivalence group is
$SO(2,2;{\bf Z})\sim SL(2,{\bf Z})\times SL(2,{\bf Z})$.
One $SL(2,{\bf Z})$ factor is nothing but the modular group
of the torus.
If we go to the dual D1-brane picture by the T-duality we have discussed,
the other $SL(2,{\bf Z})$ is also simply the modular group
of the dual torus as we will show below.

Let us introduce an orthogonal coordinate $(X,Y)$ in the D1-brane configuration
such that $Y$ represents
the direction of the T-duality.
Two vectors generating the parallelogram are
\begin{equation}
\vec v=L_x(1,\phi),\quad
\vec w=\wt L_y(\theta,1+\theta\phi).
\end{equation}
By the modular transformation, these vectors are transformed as
\begin{equation}
\vec v'=A\vec v+B\vec w,\quad
\vec w'=C\vec v+D\vec w,
\end{equation}
where $A$, $B$, $C$ and $D$ are integers satisfying $AD-BC=1$.
By this,
$L_x$, $L_y$, $\Theta$ and $\Phi$ are transformed as follows.
\begin{displaymath}
L_x'=(A+B\!\cdot\!2\pi\Theta)L_x,\quad
L_y'=(A+B\!\cdot\!2\pi\Theta)L_y,
\end{displaymath}
\begin{equation}
2\pi\Theta'=\frac{C+D\!\cdot\!2\pi\Theta}{A+B\!\cdot\!2\pi\Theta},\quad
\frac{\Phi'}{2\pi}=(A+B\!\cdot\!2\pi\Theta)^2\frac{\Phi}{2\pi}+B(A+B\!\cdot\!2\pi\Theta).
\end{equation}
These are the same with what are obtained in \cite{morita3}.

\section{Conclusions}
In this paper, we suggested the T-duality between a non-commutative
gauge theory on D2-brane with general background $2$-form field $\phi$, gauge field strength $f$ and
non-commutativity parameter $\theta$
and a theory of open strings on a D1-brane.
By using this duality,
we can represent the equivalence, which connects theories with arbitrary $\theta$'s,
and Morita equivalence
as rotations and modular transformations
of the dual configuration, respectively.

Furthermore, it may clarifies that the ambiguity of choice of parameters $\phi$ and $\theta$
comes from the freedom to choose the background $B$-field when we quantize
open strings.
Namely, when we obtain a low energy effective field theory
by quantizing open strings,
we have to fix the background field $B$.
In the dual D1-brane configuration, the direction of the T-duality is
determined by solving the equation of motion of open strings
on the corresponding background metric.
Roughly speaking, the parameter $\theta$ is determined at this point.
After the relation between the string theory and the low energy field theory
is established, the change of the $B$-field
causes the emergence of background field $\phi$ in the field theory.

\section*{Acknowledgements}
The author would like to thank W. Taylor for helpful comments.
This work was supported in part by funds provided
by the U.S. Department of Energy (D.O.E.)
under cooperative research agreement \#DE-FC02-94ER40818
and by a Grant-in-Aid for Scientific
Research from the Ministry of Education, Science, Sports and Culture
(\#9110).
%%%%%%%%%%%%%%%%%%%%%%%%%%%%%%%%%%%%%%%%%%%%%%%%%%%%%%%%%%%%%%%%%%%%%

\end{document}